\documentclass{eas}
\usepackage{graphicx}
\begin{document}
\title{Conference Summary:  Themes and Questions about the Disk-Halo Interaction}
\author{John M. Dickey}\address{University of Tasmania, Maths and Physics - Bag
37, Hobart, TAS 7005, Australia}
\begin{abstract}
The papers in this volume represent a broad spectrum of 
observational, theoretical, and computational astrophysics,
sharing as a unifying core the Disk-Halo Interaction in
the Milky Way and other spiral galaxies.  This topic covers
a wide range of Galactic and extra-galactic research, 
built on a foundation of numerous and diverse physical processes.
This summary groups the papers according to six themes, with 
some historical background and finally a look to the future.  
The final message is that the astrophysical techniques discussed
and reviewed at this conference will grow over the next decade
to answer even more fundamental questions about galaxy
evolution and the history of the universe.
\end{abstract}
\maketitle
\runningtitle{Summary: Themes and Questions}

\section{Historical Background}

This conference on the Disk-Halo Interaction in Galaxy Evolution has
been a showcase for many exciting new results in a broad spectrum of
Galactic and extragalactic research.  Later in this summary
I will briefly consider the significance of some of these results, and how
they show the way forward.  But first a few paragraphs of historical
background, for the benefit of people who are new to the field.  These
concepts were implicit in many of the papers at the conference, but
were not often mentioned explicitly.

For astrophysicists, the Disk-Halo connection is almost as old as the
interstellar medium itself.  By the early 1950's it was clear that
both cosmic rays and interstellar clouds must sometimes leave the disk.  A series
of papers culminating in Spitzer's seminal 1956 analysis
(Spitzer \cite{Spitzer_1956}, Shklovsky \cite{Shklovsky_1952}, Pickelner \cite{Pickelner_1953}), based on the gravitational potential
of the disk determined much earlier by Oort (\cite{Oort_1932}), predicted
a hot gas halo above and below the disk.  The need for cosmic
ray (CR) leakage into this halo implied that magnetic fields must connect it with
the disk.  In the decade that followed, detailed astrophysical calculations
of the heating and cooling rates of the neutral and ionized gas in the
interstellar medium (ISM) showed that thermal equilibrium between 
different phases at very different temperatures could be achieved
at a range of pressures that depends on the heavy element abundance and
the radiation field intensity (Field {\em et al.\/} \cite{Field_etal_1969}, Dalgarno
\& McCray \cite{Dalgarno_McCray_1972}).  
Meanwhile, by the early 1960's the high velocity clouds seen in 21-cm
line surveys were interpreted as evidence for both a population of
neutral clouds in the halo, and a surrounding hot gas through which
they were falling toward the disk (Oort \cite{Oort_1966}, \cite{Oort_1970}).

Dramatic observational confirmation of the existence of hot ISM gas
in the Milky Way disk came in the early 1970's from soft x-ray emission (Williamson {\em et al.\/} \cite{Williamson_1974},
Sanders {\em et al.\/} \cite{Sanders_etal_1977}) and uv absorption (Jenkins \& Meloy \cite{Jenkins_Meloy_1974}, York \cite{York_1974}).
This hot gas was immediately identified with that predicted by Spitzer for
the halo, and the cycling of gas between the disk and halo was proposed
soon after by Shapiro \& Field (\cite{Shapiro_Field_1976}, c.f. Bregman \cite{Bregman_1980}).  It was soon clear that the hot
phase gas occupies much of the disk and connects with the halo, in part
from interpretation of the anticorrelation between nearby atomic and molecular
clouds and the soft x-ray emission (McCammon {\em et al.\/} \cite{McCammon_etal_1976}, Mebold {\em et al.\/} \cite{Mebold_etal_1985}). 

The theoretical framework for understanding the hot phase of the ISM 
developed in the early 1970's as well, as the role of old supernova remnants
(SNR) in pressurizing the disk was explored quantitatively (Cox \& Smith \cite{Cox_Smith_1974}, 
Salpeter \cite{Salpeter_1976}, McKee \& Ostriker \cite{McKee_Ostriker_1977}).  By the mid-1980's it was clear
that the McKee-Ostriker paradigm was insufficient to describe the real ISM. 
The Milky Way (MW) and many other spiral galaxies have star formation rates high
enough to drive hot gas out of the disk, and with it energy in thermal
and mechanical forms (Cox \& McCammon \cite{Cox_McCammon_1986}, Cowie \cite{Cowie_1987},
Norman \& Ikeuchi \cite{Norman_Ikeuchi_1989}, Heiles \cite{Heiles_1990}), which is not included in the Mckee-Ostriker
model.  Meanwhile the power of computational
astrophysics was brought to the problem of SNR evolution (Chevalier \& Gardner \cite{Chevalier_Gardner_1974},
Mansfield \& Salpeter \cite{Mansfield_Salpeter_1974}, Struck-Marcell \& Scalo \cite{Struck-Marcell_Scalo_1984},
MacLow {\em et al.\/} \cite{MacLow_etal_1989}, Cioffi \& Shull \cite{Cioffi_Shull_1991}),
culminating in two- and finally three-dimensional simulations of the global 
ISM (Rosen \& Bregman \cite{Rosen_Bregman_1995}, Vazquez-Semadeni {\em et al.\/} \cite{VazquezSemadeni_etal_1995}, Avillez \cite{Avilles_2000}).

One of the threads running through all of these works, observational
and theoretical, is that we cannot fully understand the ISM of the MW
disk without understanding also that of the halo and how the two media interact
and exchange matter.  In the last two decades our observational data on the
ISM in the MW halo has improved dramatically.  Even more dramatic progress has
been made in studies of the ISM of other galaxies.  We can see clearly now
that galactic fountains operate in some spiral galaxies, and not in others.
We see examples of disks with infall and outflow.  It is even becoming 
possible to trace these processes as functions of redshift.  One of the
goals of this conference is to bring our detailed knowledge of the
disk-halo interaction in the MW to bear on the much larger
question of galaxy evolution over cosmic time.  

\section{Themes of This Conference}

This conference has been ambitious in its combination of topics, including
observational, theoretical, and computational studies of 
the MW disk and halo, the disks and halos of nearby galaxies, 
and gas infall and outflow in galaxies at intermediate and high
redshifts.  The papers reviewing these topics 
allow us to look back and also ahead, to see what will become possible
in the next decade or so as powerful new telescopes become available.
Rather than try to summarize the papers individually, I will consider
a few themes, and group the papers accordingly.  Of course any such grouping
is arbitrary and does not do justice to the depth and complexity of the
papers or their subjects, but there is not room for much more detail.
Citations given as names without dates refer to papers presented at
this meeting and their written versions that appear in this volume.

\section{Theme -- Magnetic Fields and Cosmic Rays Link the Disk and Halo}  

It has been understood for a long time that the ISM is
subject to a magnetic Rayleigh-Taylor instability wherein the light
fluids constituted by the magnetic field and the relativistic CR 
electrons can move through the heavier fluid constituted by the 
thermal phases of the gas to escape the gravitational potential
of the MW disk (Parker \cite{Parker_1966}).  But how much impact this
has on the dynamics of the gas in the disk and halo is still not
known.  One approach is to map the interstellar magnetic field
to see where and how the field
joins the disk with the halo.  It is possible in various ways to
map the field in the disk, as explained in the papers by Beck 
and Vall\'ee.  Searching for a vertical component of the field 
in the solar neighborhood is particularly interesting, as reported
by Mao, based on Faraday rotation surveys at high
latitudes.  Troland's paper explains how Zeeman splitting observations
at 21-cm give values for $\beta \equiv \frac{P_{kinetic}}{P_{magnetic}}$.
This quantity
is about 0.3 for the thermal pressure (with kinetic velocities given by
the temperature) but $\beta \simeq 1.2$ when non-thermal bulk velocities
are included.  This reflects the typical Mach number of about 4 for
turbulent velocities in cool, neutral medium (CNM) clouds.  Thus
the magnetic field is dynamically important in this phase, and 
probably in all other phases of the ISM, but it is not completely dominant.
It is significant, however, that magnetic pressure
is much higher than thermal gas pressure, as this could moderate
the rate of cooling.
Ferri\`ere's paper supports Troland's in these conclusions,
and gives an overview on a larger scale of the magnetic
field structure of the disk.

The dynamical importance of the magnetic field has recently been
included in simulations of the structure and dynamics of the ISM,
as reported at this conference by Breitschwerdt, Asgekar, Hanasz, and
Woltanski.  Including the CR pressure, which Breitschwerdt
couples to the gas through a spectrum of magneto-acoustic waves,
drives the gas quite effectively out of the disk, in good agreement
with the analytical prediction of Everett.  The extensive review by
Dogiel presents several fundamental equations and observations
that describe CR propagation through the disk and halo.
Some well established results from this are the scale height of
the CR population (3 kpc), the propagation path length (3 Mpc),
and hence the travel time (10$^7$ yr).  Clearly there is a lot
of energy available in the form of magnetic fields and CR's;
the CR acceleration processes alone have luminosity of more 
than 10$^{42}$ erg s$^{-1}$.  Further, per unit energy density, the
CR's are more efficient at driving mass out of the disk and
into the halo than is simple gas pressure, as Everett explains.
On the much larger scales of clusters of galaxies, the paper
by Bernet reviews the evidence for magnetic fields to at least
redshift z=1.3 as traced by Faraday rotation.  Thus we can infer that 
magnetic fields and cosmic rays have been escaping from galaxies,
at least some galaxies in some environments, for a very long time.

\section{Theme -- The Mixture of ISM Phases in the Milky Way}

The different thermal phases of the ISM are typically studied with
different spectral lines or other tracers, so understanding the
juxtaposition and relative abundances of the different phases
is like putting together separate pieces of a puzzle.  Ferri\`ere's,
Jenkins', and V\'azquez-Semadeni's papers review concepts of
pressure equilibrium, thermal equilibrium, transition of gas from
one phase to another, and how 
the mixture of phases depends on height above the plane, $z$.
The scale heights of the warm and cool phases are discussed 
further by Gaensler, Reynolds, and Kalberla.  
The ISM phases in M31 are discussed in the papers of
Berkhuijsen, Bogdan, and Braun.  The phase-mixture of the MW ISM is one of
the central topics of the conference, and the papers in this
volume represent the considered opinions of people who have been
leaders of this field for decades.  Yet the numbers presented
here, particularly the scale heights and their variation with
Galactic radius, differ in some cases by 30\% to 50\% from estimates
made a few years ago.  Clearly the diffuse ISM of the MW needs
further research!  But the papers presented at this meeting
reinforce the general consensus that we do understand the ISM
in the MW disk fairly well, and we can confidently 
extend our observing techniques and analysis tools to the halo,
to nearby galaxies, and on to high redshifts.

\section{Theme -- Structures in the MW Halo}

The scale heights of the phases do not completely describe
the structure of the ISM in the MW halo.  As Lockman puts it in
his presentation, in the halo we see ``hydrogen that has a story
to tell about how it got there.''  One of the most dramatic examples
is Smith's Cloud (Smith \cite{Smith_1963}, Lockman {\em et al.\/} \cite{Lockman_etal_2008}), a high velocity 
cloud (HVC) with mass greater than 10$^6$ M$_{\odot}$ and total 
space velocity of about 300 km s$^{-1}$, approaching the Galactic
plane at some 75 km s$^{-1}$.  This is large compared to typical
HVC's described in Wakker's review, but by no means the largest,
as Complex C is estimated to have mass 4 x 10$^7$ M$_{\odot}$.
Typical distances above or below the plane, $|z|$, are in the range 5 to 15 kpc
for these large HVC complexes.  Assuming that they represent primordial
material, their mass accretion rate averages to about 0.5 to 1 M$_{\odot}$ yr$^{-1}$.
As Putman points out in her paper, maintaining an accretion rate of
1 M$_{\odot}$ yr$^{-1}$ over the last 5 to 7 Gyr would require 
a primordial reservoir of as many as 500 dwarf galaxies with H{\tt I} mass 10$^7$ M$_{\odot}$,
or 5000 HVC complexes of 10$^6$  M$_{\odot}$.  This is
a lot, there certainly do not seem to be nearly that many
such systems in the Local Group remaining to be accreted.  
So it is important to work out the stories that the HVC complexes
have to tell, by detailed studies such as those described by
Ford, Dedes, Madsen, Haffner, Leiter, and Hill.  The larger objective
of such work is to determine what their precursors were.
Some of today's HVC complexes may have come from
dwarf galaxies that were tidally disrupted and captured by
the MW, or they may have come from tidal stripping of
an existing Local Group member, like the Magellanic Stream
discussed in the papers by Madsen, Wakker, and Putman, or perhaps they
have somehow been part of the MW halo since the epoch of galaxy formation.

There seems to be a clear distinction between the large, 21-cm selected
HVC complexes and the much lower mass clouds that dominate optical
and uv surveys toward distant halo stars, described by Wakker, Ben-Bekhti,
Nasoudi-Shoar, and Fox, and earlier by Smoker {\em et al.\/} (\cite{Smoker_etal_2004}), Lockman \& Savage (\cite{Lockman_Savage_1995}),
and see the historical data discussed by Benjamin \& Danly (\cite{Benjamin_Danly_1997}). 
The clouds seen in these surveys are similar to 
clouds seen in extragalactic absorption
studies; at least some of them are classical fountain-return clouds.
Some of the H{\tt I} HVC's are probably in this category as well.
For launching the fountain, the large Galactic chimneys and supershells
are leading contenders, as described by McClure-Griffiths, Sato, Pidopryhora,
Gon\c calves, and
Dawson.  These concentrate and direct the energy provided by many
supernova remnants to drive hot gas out of the plane and into the
lower halo.  Such structures appear to be common in nearby spiral galaxies.
The structure of the M82 superwind is reminiscent of
a galactic chimney driven to the extreme by the starburst nucleus,
as discussed by Westmoquette and Trinchieri.

\section{Theme -- HVC's in Nearby Galaxies and Groups}

One of the most successful aspects of this conference has been
the seamless merging of Galactic and extra-galactic observations
and theory.  This shows the maturity of the field, in that the
same astrophysical phenomena that we study in detail in the MW we can
trace more broadly in nearby spirals, and with fair sensitivity
even at intermediate to high redshift.  Thus many of the themes in
Dettmar's review of disk-halo interaction in nearby galaxies echo
those of Ferri\`ere's review of the same topic in the MW, and 
many other reviews include both Galactic and extra-galactic examples.
In particular, the H{\tt I} HVC's in the Local Group and nearby galaxy
groups reviewed by Putman, Oosterloo, and Chynoweth, are analogous
to the large HVC complexes in the MW halo.  But their morphology can
be completely different.  Unlike the Local Group, the M81-M82-NGC3077 group
has a neutral intra-group medium, probably a massive tidal feature,
as described by Chynowyth.  The Local Group is more typical, apparently
cases like the M82 group are rare.  But it is critically important
to get more sensitive H{\tt I} surveys of many more nearby groups, in order
to interpret the data from absorption surveys at intermediate and
high redshifts.  This will come from new arrays with enhanced 
surface brightness sensitivity, like the Australian Square Kilometre
Array Pathfinder (ASKAP).

\section{Theme -- The Halos of Spiral Galaxies seen Edge-On}

Spiral galaxies seen very nearly edge-on are excellent subjects for
study of the $z$ distribution of the ISM phases, and the structure
of the gas in ther halos.  Such studies are the topics of the papers
by Rossa, Wu, Rand, Heald, Howk, and Oosterloo's poster, and they are 
included in the reviews by Dettmar and Reynolds.  From these
studies we learn what a range of fountain activity there is in spiral galaxies,
depending on the star formation rate (SFR).  The lower halos of the edge-on
spirals that have been studied in detail show how different the
scale heights of the diffuse neutral and warm ionized media can
be, ranging from a few hundred pc to a few kpc.  Rossa also makes
the point that the morphology of the diffuse ionized gas changes with
SFR, with the high SFR galaxies showing a smooth ionized halo, and the
intermediate (lower) SFR galaxies showing many wisps that probably trace large chimneys,
bubbles, and supershells.

Another critical question that edge-on spirals can help answer is
how the rotation velocity of the gas in the halo departs from that
of the disk below.  Heald and Benjamin discuss this based on 
observations, and Fraternali reviews theoretical and computational
studies.  Here again, the SFR seems to play a critical role.  Values
for the gradient in rotation velocity with $z$ vary from -8 to -30
km s$^{-1}$ kpc$^{-1}$, with the higher values seen in the galaxies
with lower SFR's.  This may imply that high fountain activity strengthens
the dynamical connection between the disk and lower halo, essentially
acting as a viscosity that depends on SFR.  Similar numbers for the 
Milky Way are still a bit controversial; Levine {\em et al.\/} (\cite{Levine_etal_2008})
find -22$\pm$6 km s$^{-1}$ kpc$^{-1}$, see also the review by
Kalberla \& Kerp (\cite{Kalberla_Kerp_2009}).

A series of major surveys of nearby galaxies including SINGS (Kennicutt {\em et al.\/} \cite{Kennicutt_etal_2003}),
THINGS (Walter {\em et al.\/} \cite{Walter_etal_2008}), and SONG (Regan {\em et al.\/} \cite{Regan_etal_2001}) are mentioned in
a few papers in this conference, including those of Leiter and Portas.   
These surveys are primarily concerned with the disks of nearby spiral galaxies.
They are rich in information on the mixture of ISM phases, and
the links from the ISM to star formation.  They will also prove
useful for understanding the disk-halo connection.  These surveys are 
particularly interesting for revealing the structure of the outer disks, 
beyond the area of active star formation.  Conditions in such outer
disks are in some ways intermediate between those in the inner disk and
in the halo, as discussed by Kalberla and Dickey for the MW.
These surveys give a hint of the even more powerful studies of nearby
galaxies that will be possible in the coming decade with telescopes
such as ALMA, JWST, and the SKA.

\section{Theme -- Outflows from Galaxies at High Redshift}

Surveys of absorption lines in the spectra of QSO's at high redshift
are a very powerful way to trace the abundance of gas clouds
intervening along the path to the continuum source.  Studies of the
Lyman alpha forest can detect clouds with atomic hydrogen column densities
as low as 10$^{13}$ cm$^{-2}$. Ranges of progressively increasing column densities
are sampled by metal line systems, Lyman limit systems, and damped Lyman alpha systems 
finally reaching column densities of  10$^{22}$ cm$^{-2}$ or higher.
A remarkable result from forty years of such studies is that the
abundance of absorbing systems as a function of column density, $n(N_H)$, 
can be described as a power law over this entire range of
column densities, with $\frac{{\rm d}n}{{\rm d}N_H} \propto N_H^{-1.58}$, as 
reviewed by Richter.  Many of these absorption lines originate in the halos 
or outer disks of intervening galaxies; we infer that some of these clouds
 are driven by galactic fountain processes similar to those we see in
the MW and nearby galaxies.

The ISM in high redshift galaxies can also be studied using emission
line tracers, particularly at mm and sub-mm wavelengths, as reviewed
by Greve.  As ALMA comes on-line in the next few years, this field
will grow rapidly, and we may hope to detect galaxies like the MW at
redshift z$\sim$1.  For now, most of the galaxies studied at
such great distances are at the high end of the CO and far-IR luminosity
functions, typically starburst and ultra-luminous IR galaxies plus some QSO's.

With the power of 8m class optical telescopes, absorption lines can
be studied not only toward bright QSO's, but also using the light
of normal galaxies at high redshift.  The DEEP survey described by
Koo, and the specific examples presented by Rubin, show the amazing
accomplishments of high z spectroscopy at the Keck Telescope.
Both outflow and infall are seen in galaxies in the DEEP sample,
with outflow correlating with
infrared luminosity, but infall not.  Results of stacked spectra
give very useful numbers, such as that 10\%
of absorption lies beyond the escape velocities of the galaxies,
and the fact that almost all galaxies have at least some outflow
gas.  A further result is that it is not the dwarfs but the more
massive galaxies, like the MW, that dominate the total outflow 
that ultimately enriches the inter-galactic medium with heavy
elements.

To calibrate the present abundance of absorbing clouds 
vs. the numbers in the early universe requires
low redshift surveys of the far-uv lines, seen shifted into
the optical and near-IR at high redshift,
using far-uv telescopes like FUSE.  The papers
by Fox and Richter give useful comparisons
among different line tracers in the optical and uv.  An interesting
twist is the appearance of two populations of Mg II absorbers.  The
stronger class of lines is more reliably a tracer of outflows. 

Theories and questions involving galaxy evolution and the history
of disk-halo interactions were presented by Bernet, Jachym, Fangano,
Martin and Bregman.  There are some very big issues involved, such
as the ``missing baryon problem'' raised by Bregman, and the metal 
enrichment of the IGM discussed in detail by Martin.  Martin makes
the point that ``What you see is not everything that is there.''
It is very important to remember that outflows can and do happen
at temperature and density combinations for which we have no
sensitive tracers. For example, at temperature $T \sim 10^7$ K
(1 KeV x-rays) we can hardly make out the winds even from nearby
galaxies, let alone high redshift systems, as Trinchieri explains.

\section{What Next?}

Telescopes that will become available over the next
decade will bring improvements in resolution and sensitivity
of an order of magnitude or more in many wavebands.
With such improvements over current capabilities it
will become possible to study galaxies like the MW
at z=1 with the same level of detail that we now
apply to galaxies in nearby groups.  This conference
has demonstrated that {\bf extragalactic and
Galactic astrophysics merge} when the observational
data get that good.  The point is, the questions we
are asking now about nearby galaxies are the same ones
we have been asking and answering in the MW for decades.
Soon we may be doing the same for galaxies at z=1 and beyond.

Experts in this field, albeit self-appointed ones like many of us
attending this conference, stand to gain academic ground in studies of 
galaxies at high redshift, as the science shifts into
our purview.  We can start thinking now about what
we would do if we had images and spectra of galaxies
at cosmological distances as good as those we have now
for, e.g., M81 and M82.  Here are some questions 
that we might begin to answer:

\begin{itemize}
\item Do disks assemble by satellite accretion?
Why doesn't this disrupt the disk?  Does it lead to SF bursts?
If there are SF bursts, do they denude the disk?  Does
SF have a negative feedback ``governor'' that limits its speed
in a disk where the ISM can easily be lost to a wind or fountain?

\item Is the intra-cluster medium (ICM) that we see 
in nearby rich clusters
representative of the inter-galactic medium (IGM),
in metallicity and in the ratio of total baryon mass to galaxy light?  If so, how and
when did all the baryons escape from the galaxies?  Was
this happening at the
same time as the accretion that built galaxies?

\item Did the magnetic field build up slowly through a
dynamo process in galaxy disks, or was it rapidly
generated during the epoch of galaxy formation?
How much seed field came from the big bang itself?
Is there an IGM of magnetic fields and CR's similar to
those seen in the ICM of rich clusters?

\end{itemize}
In the next decade we may be confronted with the
data we need to answer these questions.  ``Confronted''
because we may not be ready to understand it all.
For want of more profound insight, the interpretation of
that data will probably follow astrophysical analysis paths
similar to those discussed in this volume as applied to
the disk and halo of the MW and nearby galaxies. 
We can look forward to extending what we have learned
about our own disk-halo connection to galaxies on much larger scales.
If this brings deeper understanding of the history of
the universe, then we can hope for more
conferences as interesting as this one in years to come.

\end{document}